# Experimental design of a millifluidic flow-focusing method
# for biomimetic nanocellulose and hemicellulose-based biopolymer fibres

A.MOISY[a], H.VOISIN[a], J.DAVY[a], B.CATHALA[a], S.GUESSASMA[a]

a. INRAE, UR1268 Biopolymères Interactions Assemblages, F-44300 Nantes, France,
amelie.moisy@inrae.fr

**Résumé :**

*Les performances mécaniques des fibres végétales sont notamment liées à la présence d'éléments cristallins dispersés au sein d'une matrice cohésive amorphe. Les propriétés mécaniques des fibres sont d'autant plus élevées que le renfort cristallin sera aligné dans l'axe de la fibre.*

*Dans le but de développer des fibres biomimétiques entièrement biosourcées, alternatives aux fibres synthétiques ou consommatrices de ressources, nous avons étudié la fabrication de filaments d'hydrogels constitués de mélanges de nanocelluloses, des nanoparticules cristallines biosourcées jouant le rôle de renfort, et de xyloglucane, une hémicellulose de la paroi végétale ayant une forte affinité pour les surfaces cellulosiques. Celles-ci assureront la cohésion entre les nanocelluloses.*

*Pour optimiser l'orientation des nanocelluloses au sein des filaments, et donc potentiellement améliorer les propriétés mécaniques des fibres, nous présentons une étude portant sur la mise au point d'une méthode millifluidique de focalisation du flux. L'idée est d'induire l'alignement des nanocelluloses en optimisant les conditions de mise en œuvre ainsi que la géométrie des circuits.*

*Le dispositif développé utilise des flux de gaines externes pour focaliser et aligner un flux central de suspension de nanocellulose. Différentes configurations en termes de concentrations, de conceptions de circuits et de vitesses d'écoulement sont testées. Des circuits imprimés en 3D sont étudiés pour produire des géométries polyvalentes et optimiser la conception du processus. Pour qualifier les orientations au cours du processus, des observations en microscopie en lumière polarisée sont effectuées, car l'alignement des structures cristallines de la nanocellulose crée une biréfringence dans les suspensions. Des déphasages optiques significatifs liés aux orientations des particules de nanocellulose sont visibles par des gradients de couleur, variant avec les concentrations et les vitesses d'écoulement des suspensions. Les essais réalisés dans le cadre de cette étude démontrent que l'orientation de la nanocellulose dans les hydrogels anisotropes a été ajustée avec succès en utilisant différentes géométries de circuits millifluidiques, avec l'introduction de xyloglucanes pour produire de nouveaux types de fibres biosourcées.*

**Abstract :**

*The mechanical performance of plant fibres is linked to the presence of crystalline elements dispersed within an amorphous cohesive matrix. The more the crystalline reinforcement is aligned with the fibre axis, the better the mechanical properties of the fibre. With the aim of developing entirely biobased biomimetic fibres as alternatives to synthetic or resource-consuming fibres, we have studied the fabrication of hydrogel filaments made from mixtures of nanocelluloses, biobased crystalline*

*nanoparticles acting as reinforcement, and xyloglucans, a plant wall hemicellulose with a strong affinity for cellulose surfaces. These will ensure cohesion between the nanocelluloses. To optimize the orientation of the nanocelluloses within the filaments, and thus potentially improve the mechanical properties of the fibres, we present a study on the development of a millifluidic method of flow-focusing. The developed setup uses external sheath flows to focus and align a nanocellulose suspension central flow. Different configurations in terms of concentrations, circuit designs and flow velocities are tested. 3D printed circuits are explored to produce versatile geometries and optimize the process design. To qualify the orientations during the process, observations with a polarized microscope (POM) are made, as the alignment of the nanocellulose crystalline structures creates birefringence in suspensions. Significant optical phase shifts related to the nanocellulose particles' orientations are visible by color gradients, varying with the suspensions' concentrations and flow velocities. Results demonstrate successful tuning of nanocellulose orientation into anisotropic hydrogels using different millifluidic circuit geometries, with the introduction of xyloglucans to produce new types of biosourced fibres.*

## 1    Introduction

Hydrogels are composed of three-dimensional polymeric networks swollen with water. They are a complex, tunable and rich family of materials used in a very wide range of domains going from biomedical to sensors or textiles, and with many different processing strategies depending on the targeted properties [1-3]. Nanocomposite hydrogels in particular are a promising way to reach superior mechanical properties [4].

Nature has developed a wide variety of hydrogel structures, and biomimicry offers an attractive approach for designing hydrogel microstructures tailored to specific applications. Our group has been working for several years on such associations, in particular on mixtures of nanocelluloses and xyloglucans [5-9]. Indeed, the cellulose and hemicelluloses self-assemble in the plant cell wall in nature, which can be reproduced with nanocelluloses such as cellulose nanocrystals (CNC) or cellulose nanofibrils (CNF) and xyloglucans (Figure 1). Depending on the parameters (volume fractions, XG / nanocellulose ratio, XG molar mass, cross-linking level…) the mixture can exhibit either a liquid or a gel behavior.

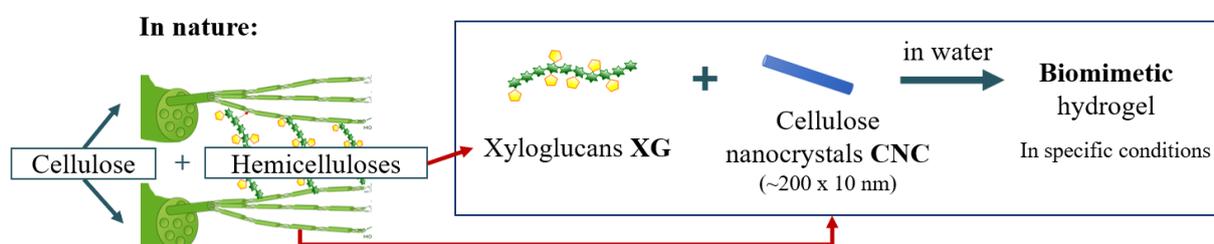

*Figure 1 Reproduction of a natural association in plants to produce a biomimetic hydrogel in specific conditions [9].*

Manufacturing biomimetic strong fibres would be relevant for applications in textile manufacturing and high-performance bio-composites. By replacing cotton or synthetic fibres by alternative sources such as wood-based polysaccharides, the environmental footprint could be reduced regarding water consumption, use of organic solvents, end-of-life treatment or energy consumption.

To optimize the mechanical performance of nanocellulose-based fibres, the principle is to favor the orientation of the nanofibres or nanocrystals in the fibre/filament direction. Different methods exist to promote this orientation: magnetic or electric fields, spinning-induced alignments, top-down strategies, extensional methods (solid or liquid state shear) [10]. The latter provides promising results, particularly with nanocellulose, with a process called flow-focusing. This method was previously applied, notably using T-shape millifluidic circuits [11-14], reaching high values of Young modulus ( until 87 Pa after drying of a hybrid nanocellulose-based fibre [12], which is comparable to glass fibres). Several parameters can play a role in the efficiency of the process: flow rates, suspensions concentrations, channel geometries.

The objective addressed in this paper is to develop an experimental nanocellulose / XG flow-focusing system with the following requirements:
- To tune the nanocellulose degree of orientation;
- To induce a self-assembly with xyloglucans and obtain biomimetic gelled filaments;
- To enable *in situ* observation;
- To allow easy production and future tuning of fibre architecture using optimization tools based on AI.

## 2 Materials and Methods

## 2.1 Materials

Two types of cellulose nanoparticles are used: cellulose nanocrystals (CNC) which are highly crystalline cellulose nanorods (200 nm long x 10 nm section) and cellulose nanofibrils (CNF) which are nanofibres of higher aspect ratio (same section for 500-2000 µm length). CNC are sourced from CelluForce (Montreal, QC, Canada) and were used as a spray-dried powder. Initially, CNC are produced from bleached Kraft pulp by sulfuric acid hydrolysis. CNF were obtained from the University of Maine (Orono, USA) and used as freeze-dried solids, with a TEMPO treatment, that specifically oxidized the C6 carbon of anhydroglucose surface groups leading to the formation of carboxylic groups on the surface of nanofibers. The xyloglucans XG, or tamarind seed gum, were purchased from DSP GOYKO FOOD & CHEMICAL (Osaka, Japan) and were used as received (dry powder), and have a molar mass equal to $8.4 \times 10^5$ g mol$^{-1}$ [15].

CNC suspensions were prepared by progressively dispersing the powder in deionized water (18.2MΩ cm resistivity, Millipore Milli-Qpurification system). XG solutions were prepared similarly by dissolving into deionized water. The solutions were left under stirring overnight at 4 °C for complete dissolution. The CNC suspensions were sonicated for 5min (2 seconds on and 2 seconds off) at 20% amplitude, with an ultrasonic probe (Q700 sonicator, 20 kHz, QSonica LLC., Newtown, CT, USA) equipped with a 12.5 mm diameter titanium microtip.

## 2.2 Flow-focusing experiment

The CNC suspensions were sonicated again before each flow-focusing test, and filtered (Millipore Millex-HV, Hydrophilic PVDF 0.45 µm), as the nanocellulose tends to agglomerate when kept still.
The first system tested, the simple flow-focusing, (Figure 2a), consists of a central flow of nanocellulose oriented by an external sheath flow of water. The interest in using a liquid sheath instead of solid walls is to avoid a temporary and turbulent movement of the nanocellulose in the vicinity of the walls and to have a more homogeneous orientation in the width of the obtained filament. Additionally, when the filament is gelled, the external sheath flow avoids circuit clogging issues.

Contrary to the fully T-shape circuits used in the literature, the central suspension goes through an internal fused silica tube (Figure 2a) and Figure 3), with an inner diameter (ID) of 700 µm, and an outer diameter (OD) of 850 µm, placing it directly in the center of the water flow, with the advantage of using only one water entrance in the circuit. The second system tested keeps the same principle, with the addition of a T-shape intersection to create a second sheath flow of XG suspension, targeting gelation along the cavity by self-assembly with the nanocellulose (Figure 2b)).

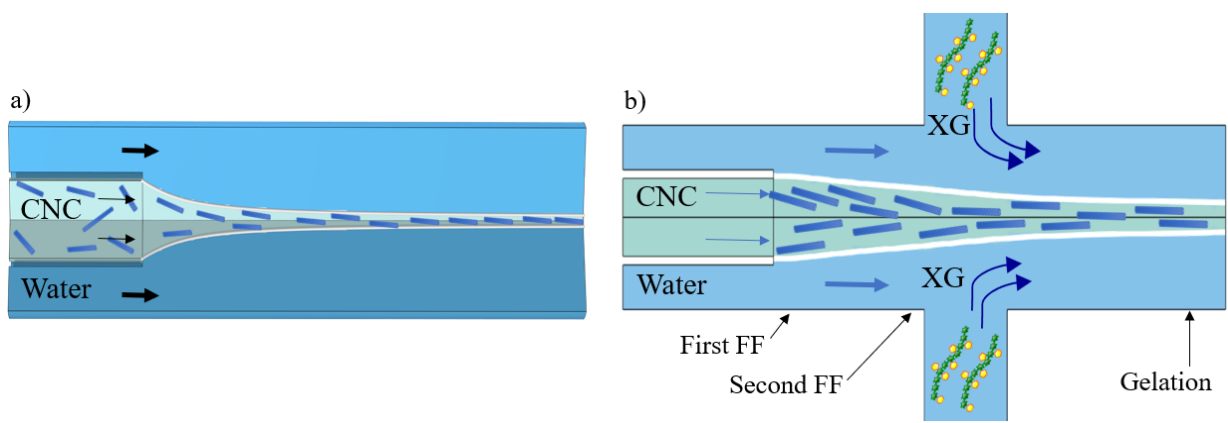

*Figure 2 a) Simple flow-focusing (FF) principle b) Double flow-focusing principle, introducing xyloglucans as second sheath flow.*

One syringe pump per liquid type is used to feed the circuit (Harvard apparatus Pump II Elite and PHD 2000, France). Regarding water and XG, 50 ml glass syringes (Hamilton) are used, whereas, for nanocellulose suspensions, a 5 ml one is sufficient as the flow rates applied are much smaller. The FEP tubes are assembled using unions and connectors (Upchurch scientific) to the flow-focusing region of the system (Figure 3a)).

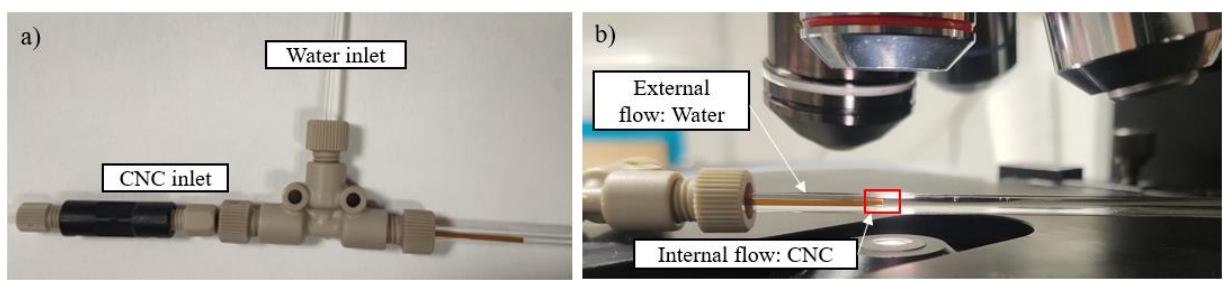

*Figure 3 Millifluidic system to create a center sheared flow of nanocellulose with an external sheath flow of water thanks to a silica tube inside a glass tube a) liquids arrival from the syringe pumps b) flows observation in the POM.*

The simple flow-focusing trials were performed in a glass tube (ID 1.62mm, OD 3mm) and the double flow-focusing trials were performed in 3D printed circuits, described below.

As the nanocellulose suspensions are increasingly birefringent with both the concentration and degree of orientation of the crystalline structures, *in situ* observation is made thanks to a Polarized Optical Microscope (POM), *i.e.* the sample is observed between polarizer and analyzer oriented at 90° one to another (Figure 3b)). The flow direction is placed at 45° of the polarizer and analyzer axis in order to maximize the birefringence intensity of the flowing suspension.

The 3D printed circuits are manufactured using the masked stereolithography (MSLA) technique, also called LCD stereolithography (LCD SLA). The principle, described in Figure 4a) and b) for the Formlabs Form 4 printer, is to photopolymerize a part layer per layer (50 µm thick layers) thanks to a liquid tank of resin above a projection screen and a moving printing platform. For the MSLA method, the projection is made thanks to LED lights (wavelength of 405 nm) and an LCD masking screen, allowing to project the shape of the layer to polymerize. Once printed, the part is washed in a convection bath of isopropyl alcohol and post-cured thanks to a UV curing system. Several trials with different machines, resins and finishing methods were made to define the best way to obtain a good visibility of the flow for *in situ* POM observation. Trials with the Anycubic system used the Photon M3 Max machine associated with the High clear resin. Tests with the Formlabs system used the Form 4 machine and the Clear resin V5. The circuits' specific geometries required manufacturing adjustments. After some trials, the printing orientation is vertical and oriented at 45° to the platform (Figure 4c)). In addition, syringes were used to clean the channels with isopropyl alcohol to complete the bath wash. Regarding the finishing, after removing the support fixtures, a polishing is performed and a spray of transparent varnish is added to the surface.

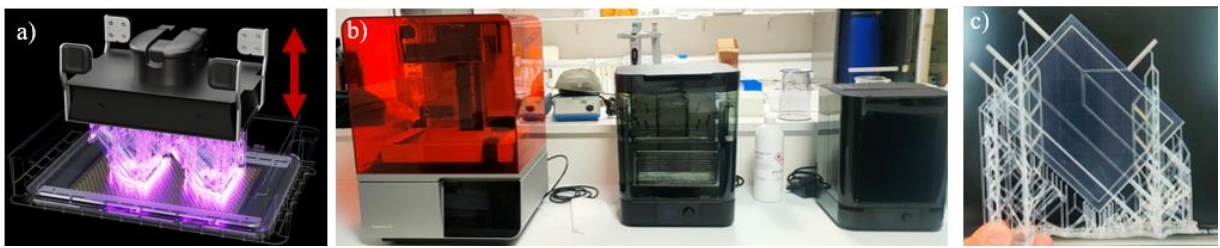

*Figure 4 a) Principle of the SLA 3D printing, with the photopolymerization slice by slice of the liquid resin in the tank (picture from formalbs.com) b) Formlabs Form 4 3D printer (left) with the washing (middle) and post-curing (right) systems c) 3D printed circuit after post-curing, before finishing steps.*

The design of various circuits is defined with Computer-Aided Design (CAD), enabling to define different cavity shapes, sizes and geometries (Figure 5a)). The circuit of Figure 4 c) was printed to evaluate the geometry options. The most suitable and convincing results were obtained with square cavities of 1.65mm width. The millifluidic assembly used for double flow-focusing results displayed in the next section is presented in Figure 5b).

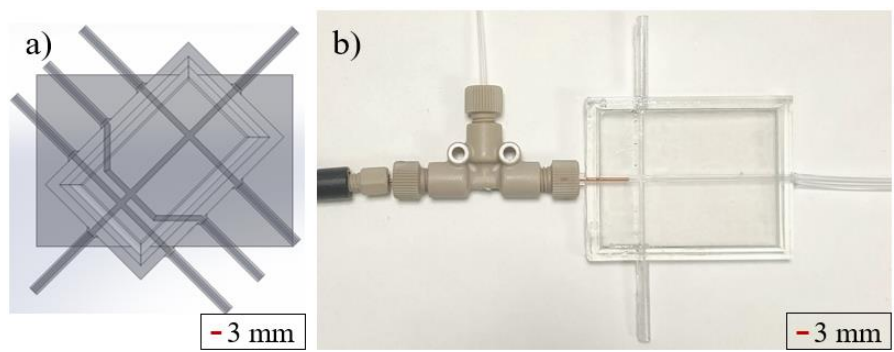

*Figure 5 a) CAO conception design of a millifluidic circuit. b) Double flow-focusing system used to introduce xyloglucans: assembly of a 3D printed circuit with square channels of 1.65 mm with the silica tube system presented above.*

# 3   Results and discussion
## 3.1   Experimental set-up design results and discussion

The initial system with a glass tube is efficient for simple flow-focusing trials, but does not allow for a second flow-focusing simultaneously with the microscopic observation. Several options were available to produce more complex millifluidic circuits. Among them, the masked stereolithography (MSLA) 3D printing method, has recently improved substantially, opening the possibility of manufacturing on command detailed transparent parts in a reasonable time. Three main challenges had then to be tackled:
- Obtain dimensionally compliant parts with narrow and long channels;
- Control the surface finishing for a sharp observation;
- Limit the birefringence of the circuit material, which can prevent a contrasted observation of the nanocellulose suspensions.

Figure 4a) displays an option produced with a first couple machine/resin (Anycubic system). Thanks to the manufacturing adjustments (orientation of the part during the print and syringe cleaning of the channels), the circuit is successfully printed with its capillaries. However, the result observations in POM exhibit high background birefringence (the channel is too bright when it should be as dark as possible) and poor sharpness. Figure 6b), by changing the machine and resin (Formlabs system), achieves better results in the channel (dark in polarized light), and improved but still insufficient sharpness. In addition, the printing time is quicker: around 3 hours with the Formlabs system against 7 hours with the Anycubic system. The resolution of MSLA 3D printers, the dimensional compliance and the properties of the printed parts are influenced by many parameters tuned differently by the equipment manufacturers: the pixel size of the LCD screen, the collimation and uniformity of the light source, the use of anti-aliasing, the minimum layer height, but also on the optical properties of the resin or the adhesion forces between layers for example [16]. Finally, the trials enabled to define the Formlabs machine and resin system as the most appropriate for this application.

To solve the sharpness issue, the most conclusive finishing method was first to polish and then varnish the circuit. Figure 6c) exposes the final result, with square channels of 1.65 mm. This manufacturing optimization enables a contrasted and sharp vision of the flow-focusing.

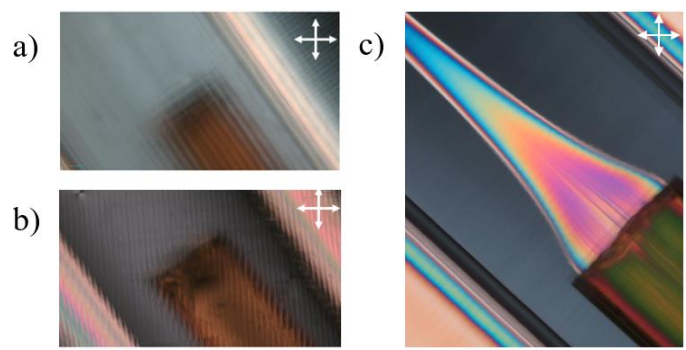

*Figure 6 a) POM visibility evaluation on a circuit printed with another machine and resin (Anycubic system) b) Visibility evaluation on the Formlabs Form 4 circuit printed with the Clear resin V5, with no finishing c) after polishing and varnishing, with a 6wt%. White arrows illustrate polarizer and analyzer orientations.*

During the trials, maintaining the 45° angle of the circuit for optimal birefringence, while observing the phenomena at different positions of the cavity was a tedious point. To facilitate this observation, the geometry of the future trials evolved to include the sides of the microscope plate linked to the circuit with the required angle. Canals and entrances were added to evaluate variations in focusing and gelation methods (Figure 7).

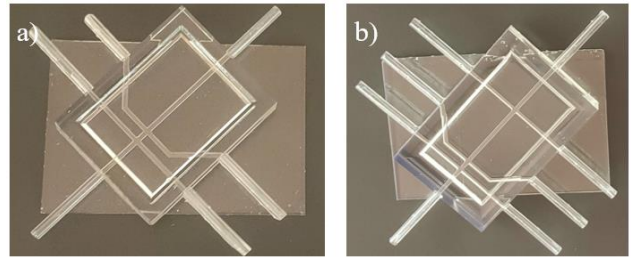

*Figure 7 Millifluidic circuits printed with integrated microscope plate and various canals geometries.*

## 3.2 Flow-focusing experiments results and discussion

Different configurations were tested in the simple flow-focusing system: CNC or CNF suspensions, at various concentrations and shear rates. The observations underline that high concentrations of nanocellulose lead to higher birefringence intensities (Figure 8). Additionally, a higher difference of flow rates between the water and the suspension, meaning higher extensional shear in the suspension, creates more birefringence. This is particularly visible for the 6 wt% CNC suspensions, displaying very organized colormaps, suggesting an orientation of the CNC along the flow direction due to extensional shear. These observations are consistent with the literature [17].

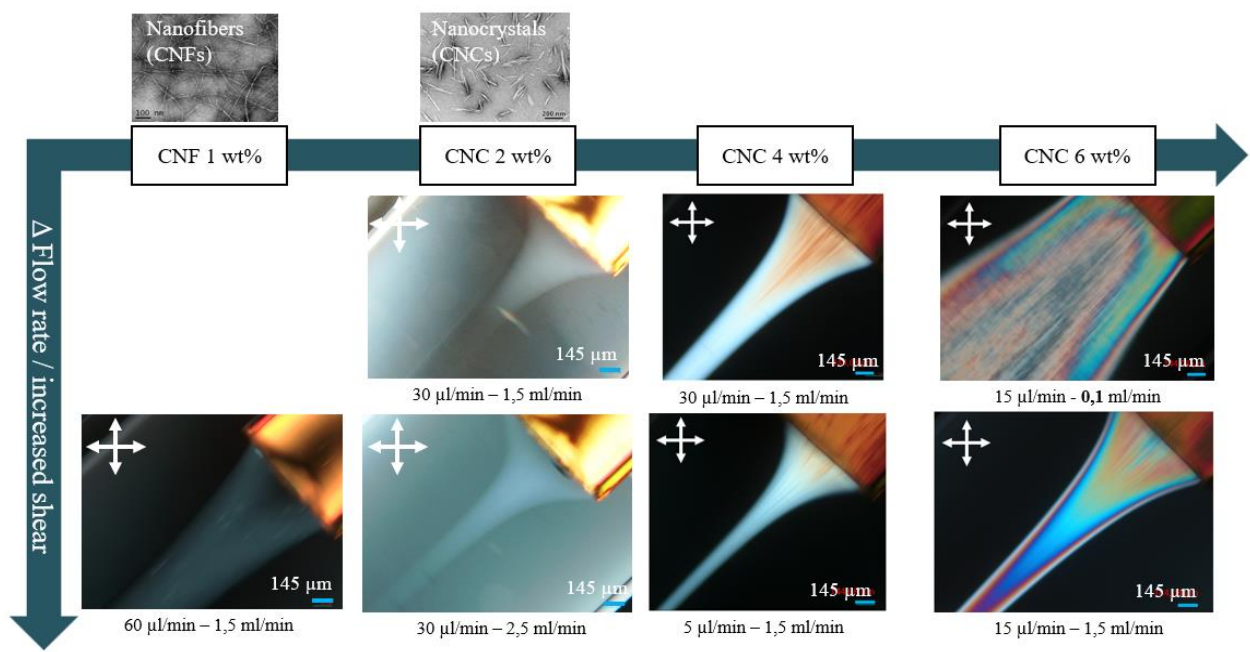

*Figure 8 Polarized optical microscope in situ pictures of the simple flow-focusing process (internal fluid: nanocellulose suspension, external fluid: water) at various concentrations of CNF and CNC and various velocities of fluids.*

The double flow-focusing results, (set-up described in Figure 2b) and Figure 5b), are presented in Figure 9. The pictures a) and b) display the exiting filament in a rotating water tank. This process was performed with a CNF (0.5 wt%) suspension central flow and an XG (0.5 wt%) solution secondary sheath flow, with blue dye for observation.

Pictures in Figure 9 c) and d) display the POM observations during a double flow-focusing with a CNC suspension (6 wt%) central flow, an XG suspension (0.5 wt%) secondary sheath flow, respectively at the second FF position (identified in Figure 2 b)) and 1 cm after the second FF position. During this test, the flow rates were 0.9 ml/min for water, 0.24 ml/min for CNC and 0.9 ml/min per side for XG.

The effect of flow-focusing is clearly visible in the 3D printed optimized circuit, and a 0.35 mm large liquid filament is obtained. The resulting birefringence colormap qualitatively correlates with the CNC degree of orientation in the extensional flow.

Once exited in the water tank, the filaments are successfully gelled, but are currently too fragile to be handled outside water without breaking. Optimization of the drying process to stabilize the filament is now the next step.

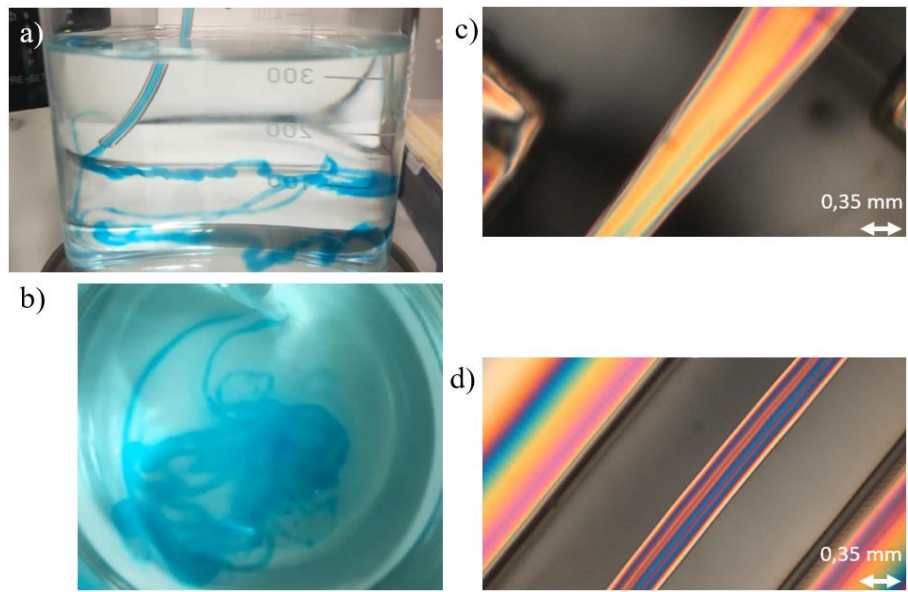

*Figure 9 Double flow-focusing results a) Exit of the system in a rotating water tank with a CNF (0.5 wt%) central flow and XG (0.5 wt%) secondary sheath flow, with blue dye for observation b) observation and handling of the obtained gelled filament c) POM image of the double flow-focusing of a CNC (6 wt%) suspension and XG (0.5 wt%) secondary sheath flow, at the second FF position d) 1 cm after the second FF position.*

## 3.3 Conclusions

The flow-focusing system developed enabled, thanks to a liquid extensional strategy, to tune the orientation of the nanocellulose, and to obtain a self-assembled biomimetic nanocellulose / XG gelled filament (0.35 mm width filaments produced in 1.65 mm width square channels). The choice of the 3D printing stereolithography technique and optimization of its manufacturing parameters enabled the *in situ* POM observation and qualitative evaluation of the nanocellulose orientation. Additionally, this circuit production method provides design freedom and ease of manufacturing compatible with future optimization trials.

The next steps of this work are to promote a stronger gelation and to initiate a drying, aiming to obtain dry biomimetic fibres. To this end, velocities, circuit geometry and suspension concentrations will be optimized for faster and stronger gelation, as well as the XG molecular weight, tunable through ultrasonic treatment [18]. The solvent of the reception tank could also be changed to promote the drying of the microstructural network. Sensitivity trials to these parameters will pave the way to obtain mechanically performant dry fibres.